\documentclass[a4paper]{article}
\usepackage{INTERSPEECH2021}
\usepackage{amsmath,graphicx,siunitx}
\usepackage[hidelinks]{hyperref}
\usepackage[acronym]{glossaries}
\usepackage{booktabs}
\usepackage{makecell}
\usepackage{tikz}
\usepackage{pgfplots}
\glsdisablehyper

\newacronym{AM}{AM}{acoustic model}
\newacronym{ASR}{ASR}{automatic speech recognition}
\newacronym{DFT}{DFT}{discrete Fourier transformation}
\newacronym{LFR}{LFR}{low frame rate}
\newacronym{nWER}{nWER}{normalized word error rate}
\newacronym{WERR}{WERR}{word error rate reduction}
\newacronym{WER}{WER}{word error rate}

\newacronym{RNN-T}{RNN-T}{recurrent neural network transducer}
\newacronym{STFT}{STFT}{short-time Fourier transformation}
\newacronym{VAD}{VAD}{voice activity detection}
\newacronym{LSTM}{LSTM}{long short-term memory}
\newacronym{CNN}{CNN}{convolutional neural network}

\title{Improving RNN-T ASR Accuracy Using Context Audio}
\name{Andreas Schwarz, Ilya Sklyar, Simon Wiesler}
\address{Amazon.com}
\email{asw@amazon.com, ilsklyar@amazon.com, wiesler@amazon.com}

\begin{document}
\maketitle
\begin{abstract}
We present a training scheme for streaming automatic speech recognition (ASR) based on recurrent neural network transducers (RNN-T) which allows the encoder network to learn to exploit context audio from a stream, using segmented or partially labeled sequences of the stream during training.  We show that the use of context audio during training and inference can lead to word error rate reductions of more than 6\% in a realistic production setting for a voice assistant ASR system. We investigate the effect of the proposed training approach on acoustically challenging data containing background speech and present data points which indicate that this approach helps the network learn both speaker and environment adaptation. To gain further insight into the ability of a long short-term memory (LSTM) based ASR encoder to exploit long-term context, we also visualize RNN-T loss gradients with respect to the input.

\end{abstract}

\section{Introduction}
\label{sec:intro}

Voice assistants like Amazon Alexa use streaming \gls{ASR} for low-latency recognition of user requests.  Streaming \gls{ASR} systems continuously process audio input without requiring ``offline'' processing of full utterances.  An example of such a system is the \gls{RNN-T} \cite{graves_sequence_2012}.

Usually, recognition for voice assistant devices is activated by a keyword (e.g., ``Alexa'') detected on device, before audio is streamed to the cloud for recognition by the \gls{ASR} system.  For efficiency reasons, the received audio stream may be decoded only partially by the \gls{ASR} system.  \gls{ASR} may be applied to segments of the stream which are defined by a keyword detector \cite{panchapagesan2016multi}, voice activity detector and endpointer \cite{maas2018combining, maas2016anchored}. E.g., the already detected keyword and/or any following silence might be skipped and decoding would be carried out on detected speech segments independently. Likewise, supervised training of such an \gls{ASR} system would use matched pairs of audio and text for each of the segments, and train on each segment independently. However, this independent handling of utterance segments has the disadvantage that acoustic context from preceding segments cannot be exploited by the \gls{ASR} system.  An alternative approach would be training \gls{ASR} on full utterances without segmentation of the audio. This approach is complicated by the fact that, in training data, not all segments of an utterance may have transcriptions available, and that it would create a mismatch with the segmented decoding approach which is desired for efficiency reasons.

In this paper, we address the problem of training an \gls{RNN-T} based streaming \gls{ASR} system for segment-wise decoding, while enabling the encoder network to learn to adapt to the environment and/or speaker by making use of the entire available acoustic context, even if, during training, only some segments have labels available. E.g., if during training the encoder sees streams containing keyword audio and a subsequent user request like ``play some music'', it could learn to focus on the speaker of the keyword part of the stream even when only decoding the second segment.

Recurrent networks such as \gls{LSTM} can in theory encode unlimited temporal context, and have been proven to be able to carry information over thousands of steps \cite{hochreiter1997lstm}. When used in hybrid deep neural network-hidden Markov model (DNN-HMM)-based \gls{ASR} systems, recurrent networks are trained using truncated backpropagation through time \cite{jaeger2002tutorial} for memory efficiency and to deliberately prevent the network from learning dependencies which are modeled by the HMM. Nowadays, this limit on the temporal context that the model can learn has mostly been lifted for the training of \gls{LSTM}-based end-to-end \gls{ASR} systems. Non-recurrent architectures such as \gls{CNN} or self-attention (transformer) \cite{zhang2020transformer} can implement arbitrarily long context at the cost of inference computation time. It has recently been shown that longer context can have significant benefits for such architectures \cite{tripathi2020transformer,han2020contextnet,hori2020transformer}. The role of sequence lengths for the training of \gls{LSTM} encoders has also been investigated in the context of the mismatch between short audio streams in training and long audio streams in inference \cite{narayanan2019recognizing,chiu2020rnnt}. In our work, we demonstrate that using additional context audio in a unidirectional \gls{LSTM} speech encoder has a significant beneficial effect on a practical \gls{ASR} task, and illustrate the capability of such an encoder to make use of such context information across several seconds of audio.

A related line of work focuses on incorporating context audio in a more explicit manner. In \cite{king2017robust,wang2019end}, speaker characteristics are extracted from an anchor segment of the utterance which is determined by a keyword detection model, and provided to the \gls{ASR} model in order allow the model to focus on the same speaker. In contrast to such approaches, we present the encoder network the entire available audio as context, and instead of explicitly defining adaptation utterances, allow the encoder network to implicitly learn to make use of available context for adaptation during training with the \gls{RNN-T} loss.

In the following, we first provide a short review of \gls{RNN-T} \gls{ASR}.  We then describe our proposed approach for training on utterances with segmented transcriptions while fully exploiting the available acoustic context.  We show experimental results which demonstrate that the proposed approach leads to \gls{WER} reductions in two systems, one trained on data selected for demonstration purposes, and one trained on a production-scale dataset. We investigate the role of learned environment and speaker adaptation in contributing to this improvement, and attempt to visualize the use of acoustic context by the \gls{LSTM} encoder.

\section{Overview of RNN-T ASR}
\label{sec:format}

We employ the \gls{RNN-T} model architecture to validate the proposed approach due to its popularity in the streaming application that we are interested in.  The \gls{RNN-T} model defines the conditional probability distribution \(P(\mathbf{y} | \mathbf{x})\) of an output label sequence \(\mathbf{y} = [y_1, \ldots, y_U]\) of length \(U\) given a sequence of $T$ feature vectors \(\mathbf{x} = [x_1, \ldots, x_T]\). The classic \gls{RNN-T} model architecture consists of three distinct modules: an encoder, a prediction network, and a joint network.  The encoder maps sequentially processed feature vectors \([x_1, \ldots, x_T]\) to high-level acoustic representations, similar to the acoustic model in the hybrid \gls{ASR} approach:
\begin{equation}
\label{eq:encoder}
\mathbf{h} = \operatorname{Enc}(\mathbf{x})\,.
\end{equation}

The prediction network (also known as decoder in the literature) takes as input a sequence of labels \( [y_1, \ldots, y_j]\). The joint network combines the output representations of the encoder and the prediction network and produces activations for each time frame \( t \) and label position \(j\), which are projected to the output probability distribution \(P(\mathbf{y} | \mathbf{x})\) via a softmax layer.

During training, the target label sequence \(  \mathbf{y}^* \)  is available and used to minimize the negative log-likelihood for a training sample:
\begin{equation}
\label{eq:rnnt-objective}
\mathcal{L}_\text{RNN-T} = - \log P(\mathbf{y}^*  | \mathbf{h})\,.
\end{equation}

In the following, we use $\mathcal{L}_\text{RNN-T}(\mathbf{h}, \mathbf{y}^* )$ to express the computation of the joint network, the prediction network, and the \gls{RNN-T} loss based on a given encoder output sequence $\mathbf{h}$ and target label sequence $\mathbf{y}^*$.

\section{Training on Segmented Data}
\label{sec:pagestyle}

We use the term \emph{utterance} to refer to the entire audio stream received by the device for one interaction of the user with the voice assistant, which typically includes both an activation keyword (``Alexa'') and the expression of the user intent (``play some music''), and has a typical length of \SIrange{3}{15}{s}.
Within an utterance, one or multiple speech \emph{segments} may be defined, e.g., by a voice activity detector \cite{maas2018combining,maas2016anchored} or by the keyword spotter.
Of these segments, only some may be selected for human transcription, either based on heuristics, e.g., excluding the activation keyword, or using a more systematic active learning approach \cite{drugman2019active}. Some of the segments may also have labels available which are not human-generated (e.g., machine transcriptions or transcriptions inferred from the detected activation keyword), while some segments may be intentionally excluded from labeling due to ambiguities (e.g., overlapping speakers or unintelligable speech).

\subsection{Baseline Training Approach}
In the baseline training approach, all labeled segments are treated independently, i.e. each segment is forwarded through the encoder separately, and the training loss for an utterance is the sum over the segment losses. Denoting the feature sequence of the $m$-th segment of an utterance as $\mathbf{x}_m = [x_{t_{\mathrm{S},m}}, ..., x_{t_{\mathrm{E},m}}]$ and the corresponding target label sequence as $\mathbf{y}^*_m$, the utterance loss is
\begin{equation}
L_\text{segmented} = \sum_{m=1}^{M}\mathcal{L}_\text{RNN-T}(\operatorname{Enc}(\mathbf{x}_m), \mathbf{y}^*_m) \,.
\end{equation}

When trained in this manner, the encoder will not be able to learn to make use of any left context $x_t$ for $t<t_{\mathrm{S},m}$ outside of the labeled segment $m$ for the decoding of that segment.
Such context could however help the encoder learn to implicitly adapt to the speaker of the activation keyword or to any characteristics of the acoustic scenario.

\begin{figure}
  \centering
      \includegraphics[width=\columnwidth]{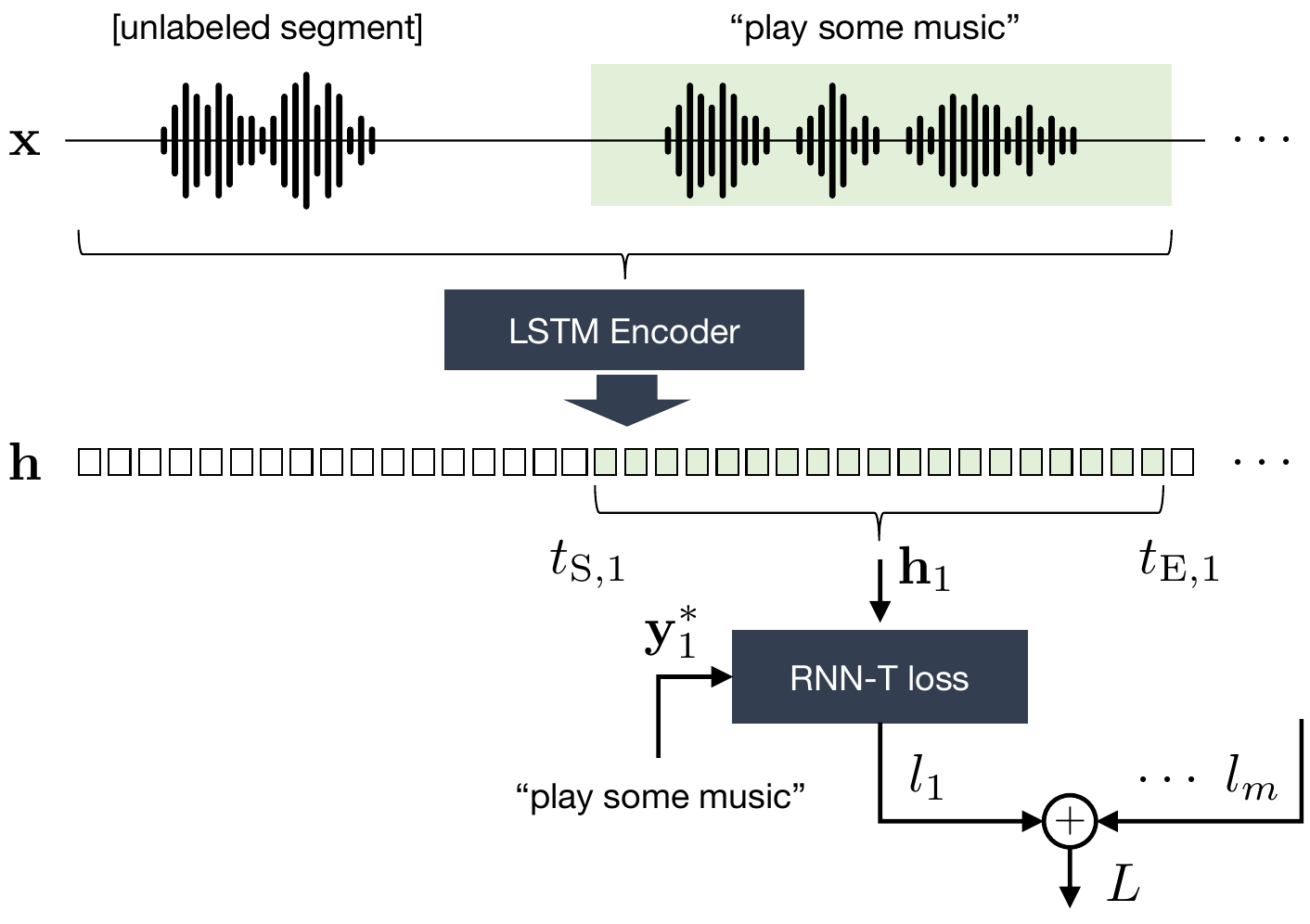}\vspace{-1mm}
  \caption{Block diagram of the RNN-T loss computation on segmented utterances.}
    \label{fig:block-diagram}
\end{figure}

\subsection{Proposed Full-Utterance Training with Context Audio}
\label{sec:training}
In the following, we describe our proposed approach for training of the encoder on the full available acoustic context, where the context may include labeled segments, but may also be completely unlabeled. The entire available feature sequence $\mathbf{x} = [x_1, x_2, ..., x_T]$ for an utterance is forwarded through the encoder to generate an encoding sequence $\mathbf{h} = [h_1, h_2, ..., h_T]$ as described in \eqref{eq:encoder}. For each labeled segment $m$ with start index $t_{\mathrm{S},m}$ and end index $t_{\mathrm{E},m}$, we then extract the corresponding encoding subsequence:
\begin{equation}
\label{eq:encoding-subsequence}
\mathbf{h}_m = [h_{t_{\mathrm{S},m}}, ..., h_{t_{\mathrm{E},m}}].
\end{equation}
With the segment target label sequence $\mathbf{y}^*_m$ we proceed to calculate the \gls{RNN-T} loss for the segment:
\begin{equation}
l_m = \mathcal{L}_\text{RNN-T} \left( \mathbf{h}_m, \mathbf{y}^*_m \right),
\end{equation}
where $\mathcal{L}_\text{RNN-T}$ comprises the prediction network, joint network, and loss computation from a given encoding and label sequence.
Since $\mathbf{h}_m$ depends on the entire input sequence up to $t_{\mathrm{E},m}$, this loss corresponds to the negative log probability of the $m$-th label sequence given the entire input sequence until $t_{\mathrm{E},m}$, i.e., $-\log P(\mathbf{y}^*_m|\mathbf{x}_{1 \dots t_{\mathrm{E},m}})$.

The overall loss for the utterance is given by the sum of the $M$ segment losses:
\begin{equation}
\label{eq:sum_of_segment_losses}
L_\text{full-utterance} = \sum_{m=1}^{M} l_m.
\end{equation}
The optimization goal is therefore the maximization of the probability of the label sequences of all labeled segments, given the entire input sequence up to the end of each respective segment.

For the training of the model we compute the gradient of this combined loss with respect to the encoder, prediction network, and joint network parameters, backpropagating all the way through the input feature sequence.
This allows us to optimize the model for the recognition of the labeled sequences given the entire available input audio, while not restricting the encoder output for time indices where no label information is available, as illustrated in \autoref{fig:block-diagram}. %

As a possible extension, to allow for segment labels with different uncertainties (e.g., human- and machine-generated labels), we could apply weights to the segment losses accordingly in \eqref{eq:sum_of_segment_losses}.

We note that the proposed training process affects only the context seen by the encoder network, not the prediction network. While label sequences of multiple segments will in practice not be independent, we do not consider this aspect in the scope of this paper, but focus on the effect of context exploitation by the encoder only.

\subsection{Inference with Context Audio}

To exploit context audio during inference, we apply the encoder to the entire feature sequence of the utterance, segment the encoder output according to the requirements of the application (e.g., skipping the activation keyword segment), followed by standard \gls{RNN-T} beam search decoding on each segment of the encoder output as given by \eqref{eq:encoding-subsequence}.%

\section{Experimental Setup}
\label{sec:experiments}

We evaluate the effect of the proposed full-utterance training, as opposed to training on segmented audio, on two systems. System 1 is a reduced-size model and trained on an artificially reduced dataset. System 2 is a production-scale model trained on a production-scale dataset.

\subsection{Model Configuration}

Our \gls{RNN-T} system consists of a unidirectional \gls{LSTM} \cite{hochreiter1997lstm} encoder, \gls{LSTM} prediction network and feed-forward joint network.
The encoder for System 1 uses 5x1024 \gls{LSTM} layers, while in system 2 it is enlarged to 8x1024 \gls{LSTM} layers.
System 1 uses a 2x1024 \gls{LSTM} prediction network and one feedforward layer with 512 units and tanh activation in the joint network, followed by a softmax layer with an output vocabulary size of 4000 wordpieces.
System 2 uses the same prediction network architecture, but the size of the feed-forward layer in the joint network is increased to 1024 units. \autoref{table:models} summarizes the model and training hyperparameters of both systems.

\begin{table}[b]
    \caption{Model and training hyperparameters for both system configurations.}
    \label{table:models}
    \footnotesize
    \vspace{-5mm}
    \begin{center}
        \begin{tabular}{ l r r r r }
            \toprule
            & \multicolumn{2}{c}{System 1} & \multicolumn{2}{c}{System 2} \\
            & Segm. & Full utt. & Segm. & Full utt. \\
            \midrule
            Encoder              & 5x1024    & 5x1024    & 8x1024   & 8x1024     \\
            Prediction network              & 2x1024    & 2x1024    & 2x1024   & 2x1024     \\
            Joint                & 1x512& 1x512& 1x1024 & 1x1024\\
            \# Output units          & 4000      & 4000       & 4000 & 4000 \\
            \# Params            & 58M       & 58M       & 89M & 89M \\
            Batch size           & 3200      & 3200      & 1536 & 1536 \\
            Iterations           & 250k      &   250k    & 580k & 580k  \\
            Labeled audio [h]    & 14k       & 14k       & 38k & 38k \\
            Context audio [h]    & 0       & 11k       & 0 & 10k \\
            \midrule
        \end{tabular}
    \end{center}
    \vspace{-4mm}
\end{table}

\subsection{Data and Training}
\label{sec:data-and-training}
All experiments are performed on an internal dataset of de-identified recordings from voice-controlled far-field devices.
For the training of System 1, we select a subset of utterances which have at least two segments. For demonstration purposes, we discard the transcription of the first segment (typically the activation keyword, e.g., ``Alexa'') if such a transcription is available. The training dataset of System 2 has been created without such filtering and is therefore more representative of a production dataset. Approx. \SI{45}{\percent} of the utterances consist of two or more segments, with typically only one of these segments labeled. \autoref{table:models} gives the overall size of labeled audio and context audio in the training datasets. Context audio is unlabeled and may contain speech, silence or background noise.

For each system, we train a baseline variant using standard \gls{RNN-T} loss on segmented labeled audio (1a/2a), and a variant where the encoder processes the full utterance audio including context audio as proposed in \autoref{sec:training} (1b/2b).

For both System 1a and System 1b we use a training batch size of 3200 segments (1a) or utterances (1b). For System 2a and 2b, which have a larger encoder network, we use a smaller batch size of 1536 segments (2a) or utterances (2b).
Since each utterance typically contains only one labeled segment, the number of label sequences seen per batch is approximately the same for both training variants, while the amount of audio frames passed to the encoder input for the full-utterance-trained model is increased by the relative amount of context audio included in the training (\SI{78}{\percent} for System 1 or \SI{26}{\percent} for System 2).

We train using the Adam optimizer with a warm-up, hold, and exponential learning rate decay policy for a total of 250k and 580k iterations for System 1 and System 2, respectively, and select the best out of six models from the last 30k training iterations by decoding on a development dataset.  %

The acoustic features are 64-dimensional Log-Mel-Frequency features with a frame shift of 10ms, stacked and downsampled by a factor of 3, corresponding to an encoder frame rate of \SI{30}{ms}.
We use an adaptive variant of a feature-based augmentation method, SpecAugment \cite{Park2019}, as proposed in \cite{Park_2020}.
We apply two frequency masks with a maximum size of 24 in all experiments.
Time mask size and multiplicity is adapted to the length of the audio signal to ensure that time masking aggressiveness is consistent for both segmented and full utterance training.

During decoding we use a matched process with training for the forwarding of the encoder, where we segment the audio before forwarding for System 1a/2a, and forward the encoder on the full utterance for System 1b/2b. We then perform beam search with a width of 16, using the encoder sequences corresponding to the start and end timestamps of the segments.

\section{Results}

We evaluate on two different datasets for System 1 and 2, each of which is matched to the respective training setup.
For System 1, we use a filtered dataset as described in \autoref{sec:data-and-training}, while for System 2, we evaluate on a test set which is representative of production data.  For System 1 and 2, we report \gls{nWER}, where a value of 1.00 corresponds to the performance of System 1a or 2a on the overall test set, respectively. We also report the relative \gls{WERR} obtained with the system trained using our proposed method compared to the baseline.

\autoref{table:wer} summarizes the results. For each system we report results on the overall evaluation dataset, as well as results on subsets containing utterances with only foreground speech (``clean''), utterances with interfering background speech within the labeled segment, and utterances with speaker changes within the utterance (e.g., the speaker of the activation word is different from the speaker expressing the intent). We can see that our proposed method for training on full utterances leads to a significant WER reduction in both setups, especially for audio with background speech.  We hypothesize that this improvement stems from the \gls{RNN-T} encoder learning to implicitly adapt to the speaker and/or the environment within the utterance based on the additional left context seen by the encoder.  Also, we observed during training that the system trained on full utterances performs consistently better on the development set, while having a larger training loss, indicating that longer encoder input sequences have a regularizing effect.

\begin{table}[]
    \caption{Comparison of systems with encoder trained on segmented audio vs. on full utterances, showing normalized WER (nWER) and relative WER reduction (WERR) by full-utterance training.}
    \label{table:wer}
    \vspace{-5mm}
    \begin{center}
        \begin{tabular}{ l c c r }
            \toprule
            & \multicolumn{2}{c}{System 1} & \\
                        & 1a (segm.) & 1b (full utt.) & WERR\\
            \midrule
            \textbf{Overall test set}     & 1.00      &  \textbf{0.94}     & 6.4\% \\
            clean                & 0.92      &  \textbf{0.86}     & 6.0\% \\
            w. background speech & 1.26      &  \textbf{1.17}     & 7.1\% \\
            	w. speaker change    & 3.17		 &  \textbf{3.15}	 & 0.4\% \\
            \midrule
            \midrule
            & \multicolumn{2}{c}{System 2} & \\
                        & 2a (segm.) & 2b (full utt.) & WERR\\
            \midrule
            \textbf{Overall test set}     & 1.00      & \textbf{0.94}      & 5.8\% \\
            clean                & 0.89      & \textbf{0.85}      & 5.1\% \\
            w. background speech & 1.30      & \textbf{1.21}      & 7.1\% \\
			\bottomrule
        \end{tabular}
    \end{center}
    \vspace{-4mm}
\end{table}

\subsection{Speaker and Environment Adaptivity}
To investigate the role of speaker adaptation, we evaluated on a subset of the data which has been annotated to contain speaker changes during the utterance (\autoref{table:wer}).
This subset is generally much more challenging, due to the presence of multiple (sometimes overlapping) speakers who are addressing the voice assistant. On this dataset, the proposed model trained on full utterances does not achieve a significant improvement, indicating that the overall improvement seen on the overall test set is partially related to the model being able to adapt to one speaker.

\begin{table}[b]
    \caption{nWER of System 1a/1b on original test set, and with additional artificial reverberation of the full utterance vs. on only the decoded segment of the utterance.}
    \label{table:artificial-reverb}
    \vspace{-5mm}
    \begin{center}
        \begin{tabular}{ lccr }
            \toprule
            & \makecell{1a\\(segm.)} & \makecell{1b\\(full  utt.)} & WERR \\
           \midrule
            Base test set         & 1.00      & \textbf{0.94}     & 6.4\% \\
            +reverb on full utterance         & 1.85      & \textbf{1.51}     & 18.4\% \\
            +reverb on decoded segm.      & \textbf{1.85}    &  2.03    & -9.6\% \\
            \bottomrule
        \end{tabular}
    \end{center}
\end{table}

To also investigate the role of environment adaptation, we conduct an experiment where we apply artificial reverberation on either the full utterance or only the audio corresponding to the decoded segments of the utterance.
Reverberation is implemented by convolving with an impulse response which is randomly drawn from a database of measured room impulse responses, and re-normalizing the signal to the original power.
From the results in \autoref{table:artificial-reverb} we first observe that the gain of the model trained on full utterances over the model trained on segments is significantly higher on the artificially reverberated test set (18.4\% WERR) compared to the overall test set (6.4\% WERR), indicating that training with context audio is particularly beneficial under challenging acoustic conditions.
We also observe that, in the case where we artificially introduce an environment mismatch between the decoded segment and the rest of the utterance, the model trained on full utterances degrades \gls{WER} by 9.6\%.
We take this as an indication that the improvement seen by full-utterance training stems partially from the model learning implicitly to adapt to the acoustic environment.

\subsection{Illustration of Gradients}

The improvement seen by training on full utterances indicates that the \gls{LSTM} encoder is capable of exploiting long-range dependencies over several seconds.  As an attempt to better understand this behavior, we visualize the dependency of the loss computed with a converged model (System 1b) on past input data by computing the gradient of the \gls{RNN-T} loss with respect to the input feature vector $x_t$, i.e., $\partial L/\partial x_t$.  We show the L2 norm of this gradient for each time frame, as well as the input feature energy distribution, for an example utterance in \autoref{fig:gradients}. This example utterance consists of three segments separated by relatively long pauses. The loss in this example is computed only for the highlighted segment containing the spoken words ``play some music''.  We can observe that, while the first activation keyword occurs approx.\,six seconds before the segment for which the loss is computed, it still contributes significantly to the loss.
We consider it noteworthy here that input data can contribute to the loss even after more than a hundred steps through the recurrent \gls{LSTM} encoder.

\pgfplotsset{every tick label/.append style={font=\scriptsize}}

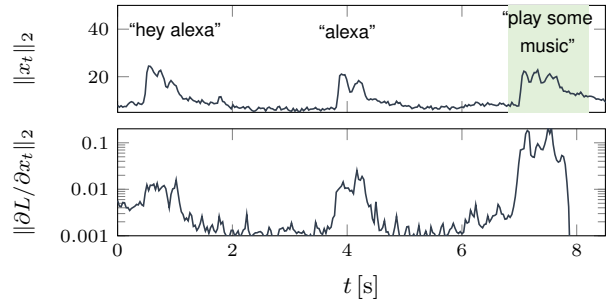
\begin{figure}
\pgfplotsset{scaled y ticks=false}
\begin{tikzpicture}
	\definecolor{clr1}{RGB}{51,63,79}
	\definecolor{clr2}{RGB}{226,240,218}
	\begin{axis}[height=3cm,width=\columnwidth,name=first,xmin=0,
		xticklabel=\empty,
		ylabel=$\|x_t\|_2$,
		ylabel style={font=\footnotesize},
		ymax=50,
		ymin=5,
		xmax=8.5
	]
	\fill[clr2] (axis cs:6.8,0) rectangle (axis cs:8.2,100);
	\addplot [mark=none, color=clr1,semithick] table [x=t, y=feat, col sep=comma] {figures/gradient.csv};
	\node[] at (axis cs: 1.0,39) {\scriptsize{\textsf{``hey alexa''}}};
	\node[] at (axis cs: 4.0,39) {\scriptsize{\textsf{``alexa''}}};
	\node[align=center] at (axis cs: 7.49,39) {\scriptsize\textsf{``play some}\\\scriptsize\textsf{music''}};
	\end{axis}
	\begin{axis}[height=3cm,width=\columnwidth,xlabel={$t\,\mathrm{[s]}$},
		at=(first.below south west),
	    anchor=north west,
    	yshift=0,
		name=second,
		xmin=0,
		ylabel=$\|\partial L/\partial x_t\|_2$,
		ylabel style={font=\footnotesize},
		ymin=0.001,
		ymax=0.2,
		xmax=8.5,
		yticklabel style={
	        /pgf/number format/fixed,
	        /pgf/number format/precision=5
		},
		scaled y ticks=false,
		ymode=log,
		log ticks with fixed point,
		x label style={at={(axis description cs:0.5,-0.3)},anchor=south}
	]
	\addplot [mark=none, color=clr1,semithick] table [x=t, y=grad, col sep=comma] {figures/gradient.csv};
	\end{axis}
\end{tikzpicture}\vspace{-2mm}
	\caption{Illustration of input feature energy distribution (top) and the gradient of the loss with respect to the input features for an example utterance.  The segment used for the loss computation (``play some music'') is highlighted.}
	\label{fig:gradients}
\end{figure}

\section{Conclusion}
\label{sec:refs}

We have proposed an approach for training an \gls{RNN-T} \gls{ASR} system using the full audio stream of an utterance as input to the encoder, while computing the training loss on one or multiple segments of the utterance.  We have shown that, in a segmented decoding setup, this approach can lead to a significant reduction in WER due to the exploitation of context audio, without necessarily requiring labels for this context.  We found indication of the model learning to implicitly adapt to the speaker and environment during the utterance, which provides a possible explanation for the observed improvement.  Furthermore, we demonstrated that a unidirectional \gls{LSTM} speech encoder network can learn to exploit long-range dependencies over more than a hundred recurrent iterations.  
\vfill\pagebreak

\bibliographystyle{IEEEtran}
\bibliography{stream_training}

\end{document}